\documentclass[seceq]{ptptex}

\usepackage{graphicx}

\newcommand{\Sfl}[1]{\ooalign{\hfil/\hfil\crcr $#1$}}

\notypesetlogo                       

\title{Quasi-Quark Spectrum in the Chiral Symmetric Phase\\ 
from the Schwinger-Dyson Equation
}

\author{
Masayasu \textsc{Harada}$^{},$\footnote{E-mail: harada@hken.phys.nagoya-u.ac.jp} 
Yukio \textsc{Nemoto}$^{}$\footnote{E-mail: nemoto@hken.phys.nagoya-u.ac.jp}
and Shunji \textsc{Yoshimoto}$^{}$\footnote{E-mail: yoshimoto@hken.phys.nagoya-u.ac.jp} 
}

\inst{
Department of Physics, Nagoya University, Nagoya 464-8602, Japan
}

\abst{%
We non-perturbatively study the fermion spectrum 
in the chiral symmetric phase from the Schwinger-Dyson equation
with the Feynman gauge, 
in which we perform an analytic continuation of the solution on the 
imaginary time axis to the real time axis with a method 
employing an integral equation. 
It is shown that the fermion spectrum has two peaks, 
which correspond to the normal quasi-fermion and the plasmino, 
although these peaks in the strong coupling region are very broad, 
owing to multiple scatterings with gauge bosons. 
We find that the thermal mass of the quasi-fermion saturates at some value of
the gauge coupling, beyond which the thermal (pole) mass satisfies $M \sim T$, 
independently of the value of the gauge coupling. 
We also comment on the appearance of a three-peak structure in the 
fermion spectrum as a non-perturbative effect.
}

\begin{document}

\maketitle

\section{Introduction}
\label{sec:intro}
Recent experimental studies carried out at the Relativistic Heavy Ion 
Collider (RHIC) have yielded unexpected findings\cite{Arsene:2004fa}.
Among them, it has been found that the collective flow of the created
matter behaves like a perfect fluid.
This suggests that the quark-gluon plasma (QGP) near the phase
transition is a strongly interacting system.
This result seems to be consistent with the findings of recent studies
employing lattice QCD,
which show that the lowest charmonium state survives at temperatures ($T$)
 higher than the critical temperature ($T_c$)\cite{Asakawa:2003re}.
Theoretically, the existence of hadronic states in QGP was predicted 
many years ago on the basis of symmetry arguments of the chiral phase 
transition\cite{Hatsuda:1985eb}.
There are also many recent studies investigating possible hadronic bound
states in QGP which were motivated by the RHIC
experiments\cite{Shuryak:2004cy,Brown:2003km}. 

The study of quarks and gluons, as well as hadronic states, 
is also important, because they are liberated degrees of freedom in QGP.
The rapid change of the energy density around $T_c$ obtained in lattice QCD
suggests that quarks and gluons actually come into play in thermodynamics.
However, in general, particles in medium have different spectra from 
those in vacuum, and thus it is quite nontrivial to determine whether 
the quasi-particle picture of quarks and gluons remains valid in strongly
coupled matter, such as that described by QGP near $T_c$. 
It is shown in Ref.~\citen{Schaefer:1998wd} that 
the quark spectrum near $T_c$ has two massive modes
with thermal (pole) masses of the order of $T$, 
taking the gluon condensate into account.
In Ref.~\citen{Petreczky:2001yp}, 
the dispersion relations of the quark and the gluon above $T_c$ of 
the deconfinement transition are computed 
in quenched lattice QCD using the maximum entropy method.
The results of those works are that the thermal masses of the quasi-quark and
the quasi-gluons are larger than $T$ near $T_c$, 
while no pure collective modes like the plasmon or the plasmino were found.
Recently, the thermal mass of the quark was also studied in quenched lattice
QCD with two-pole fitting for the spectral function\cite{Karsch:2007wc}. 
In that work, it was found that the thermal mass is of the order of $T$ 
near $T_c$.
In Ref.~\citen{Mannarelli:2005pz}, 
with a Brueckner-type many-body scheme and data from the heavy quark potential
in lattice QCD, both the thermal mass and the width of the quasi-quark are
found to be of the order of $T$ for $T=1$--$2 T_c$.
In Ref.~\citen{Kitazawa:2005mp}, it is shown in the Nambu--Jona-Lasinio
model that the quark spectrum near $T_c$ has three peaks, 
which result from coupling with fluctuations of the chiral
condensate, and that the thermal masses of the normal quasi-quark and the
plasmino are of the order of $T$.

In this paper, using a gauge theory, we investigate the fermion spectrum in
the chiral symmetric phase. 
In the weak coupling region at finite $T$, 
the HTL resummation has been established, and the fermion spectrum is well
understood at leading order\cite{Klimov:1981ka}.
In this study, we compute the fermion spectrum 
over a wide range of value of the coupling constant
using the Schwinger-Dyson equation (SDE) with the Feynman gauge. 
The SDE used in this paper includes the HTL of the fermion self-energy at 
leading oder, and it thus reproduces the fermion spectrum of HTL in the weak
coupling region. 
In the strong coupling region, by contrast, the SDE incorporates
non-perturbative corrections as an infinite summation of a certain kind of
diagrams. 
In addition, one of the advantages of our approach is that the SDE
respects the chiral symmetry and describes dynamical breaking. 
This contrasts with the fact that it is difficult to respect the chiral
symmetry on a lattice. 
Furthermore, it is straightforward to extend our formulation to finite density.

The contents of this paper are as follows.
In \S\ref{sec:SDE}, 
we formulate the SDE in the imaginary time formalism
with the ladder approximation. 
We perform an analytic continuation of the fermion propagator 
to the real time axis numerically
with a method employing an integral equation\cite{ac}.
In \S\ref{sec:SDE-sp},
using the retarded fermion propagator obtained in this way,
we investigate the fermion spectrum above $T_c$, focusing on dependences
on the gauge coupling.
Section \ref{sec:sum} is devoted to a brief summary and discussion. 
In the appendices, the notation and some technical details of the
calculations given in the text are presented.

\section{Schwinger-Dyson equation at finite temperature}
\label{sec:SDE}
In this section, we introduce the SDE with the ladder approximation at
finite $T$. 
The SDE has been employed in analyses of strong coupling 
gauge theories in vacuum\cite{kugo} and in the determination of the 
phase structure of the chiral and color superconducting
phase transitions at finite $T$ and/or 
density\cite{Harada:1998zq,Ikeda:2001vc}.

In the imaginary time formalism, the SDE
for the full fermion propagator $\mathcal{S}(i\omega_n,\vec{p})$ 
is given by 
\vspace{4em}
\begin{align}
  &\mathcal{S}^{-1}(i\omega_n,\vec{p})-\mathcal{S}^{-1}_{\rm{free}}
  (i\omega_n,\vec{p})
  \notag \\
  &= - T\sum_{m=-\infty}^{\infty} 
  \int\frac{d^3k}{(2\pi)^3} 
  g^2\gamma_{\mu}\mathcal{S}(i\omega_m,\vec{k})
   \Gamma_{\nu} 
  \mathcal{D}^{\mu\nu}(i\omega_n-i\omega_m,\vec{p}-\vec{k})
   \,\,,\,\,
\label{eq:fullSDE}
\end{align}
where $\Gamma_{\nu}$ and $\mathcal{D}^{\mu\nu}$are the full 
fermion-gauge boson vertex function and the Matsubara propagator for 
the gauge boson, respectively, and $\omega_n=(2n+1)\pi T$ is the Matsubara
frequency of the fermion.  
(For the definition of the Matsubara Green function, see Appendix
\ref{app:note}.) 
The quantity $g$ is the fixed gauge coupling. 
In this study, the current fermion mass is taken to be zero,
and thus the free fermion propagator is given by
$\mathcal{S}_{\rm free}^{-1}=-\Sfl{p}=\vec{p}\cdot \vec{\gamma}-i\omega_n 
\gamma_0 $. 
Here, $\mathcal S$ is restricted by rotational invariance in space and parity
invariance to take the form 
\begin{align}
\mathcal{S}(i\omega_n,\vec{p})=
 \frac{1}{B(i\omega_n,p)+A(i\omega_n,p)\vec{p}\cdot\vec{\gamma}-C(i\omega_n,p)i\omega_n\gamma_0}
 \,\,,
\label{eq:Sm}
\end{align}
with $p=|\vec{p}|$.
Equation (\ref{eq:fullSDE}) is not closed for the full fermion propagator. 
Here, we use the ladder approximation in which 
$\mathcal{D}^{\mu\nu}$ and $\Gamma_{\nu}$ are replaced with the 
tree-level quantities, $\mathcal{D}^{\mu\nu}_{\rm free}$ and $\gamma_{\nu}$,
respectively.
Then, Eq.~(\ref{eq:fullSDE}) becomes a closed equation for $\mathcal{S}$, 
\begin{align}
&\mathcal{S}^{-1}(i\omega_n,\vec{p})-\mathcal{S}^{-1}_{\rm{free}}
 (i\omega_n,\vec{p}) 
\notag \\
&= - g^2 T\sum_{m=-\infty}^{\infty}\int\frac{d^3k}{(2\pi)^3} 
  \gamma_{\mu}\mathcal{S}(i\omega_m,\vec{k})
  \gamma_{\nu} 
  \mathcal{D}^{\mu\nu}_{\rm{free}}
  (i\omega_n-i\omega_m,\vec{p}-\vec{k}) \,\,,
\label{eq:ladderSDE}
\end{align}
where the free gauge boson propagator $\mathcal{D}^{\mu\nu}_{\rm{free}}$ 
is given by
\begin{align}
\mathcal{D}^{\mu\nu}_{\rm{free}}(i\omega_n-i\omega_m,\vec{p}-\vec{k})=\frac{1}{l^2}\left(g^{\mu\nu}-(1-\alpha)\frac{l^{\mu}l^{\nu}}{l^2}\right) \,\,,
\end{align}
with $l^{\mu}\equiv p^{\mu}-k^{\mu}=(i\omega_n-i\omega_m,\vec{p}-\vec{k})$ and 
$\alpha$ being the gauge parameter.
Here, we should mention the gauge fixing
of the SDE with the ladder approximation. 
The ladder approximation, expressed by $\Gamma_{\nu}=\gamma_{\nu}$, implies 
that the vertex
renormalization factor is unity, i.e. $Z_1=1$. 
In QED, the Ward-Takahashi identity implies $Z_1=Z_2$, where
$Z_2$ is the wave function renormalization factor for the fermion.
To satisfy this equality, $Z_2$ must also be unity.
At $T=0$, it can be shown that $A(p_0,\vec{p})=C(p_0,\vec{p})=1$, 
and thus $Z_2=1$ is always satisfied if the Landau gauge is adopted 
\cite{Georgi:1989cd}.
For this reason, 
the Landau gauge fixing is often adopted at
$T=0$ to study chiral dynamics in QCD.
At finite $T$, however, 
both $A$ and $C$ deviate from unity, owing to thermal effects, 
even in the Landau gauge\cite{Ikeda:2001vc}.   
Therefore, there is little advantage of the Landau gauge at finite
$T$. 
Here, we use the Feynman gauge,\footnote{
In Ref.~\citen{Nakkagawa:2007hu}, 
a non-linear gauge is formulated to satisfy $Z_1=Z_2$ 
in the SDE with the ladder approximation at finite $T$.}
in which the analytic continuation
of the fermion propagator is simple, as discussed in the 
following section.

By making suitable projections, the SDE (\ref{eq:ladderSDE}) 
can be divided into three coupled equations for the functions
$B(i\omega_n,p)$, $A(i\omega_n,p)$, and $C(i\omega_n,p)$.
After performing the three-dimensional angle integral, 
they are given by
\begin{align}
 B(i\omega_n,p) &=  2 \pi T \sum_{m = -N}^{N} \int_{0}^{\Lambda} dk  
 \frac{k^2 B(i\omega_m,k)E_1(i\omega_n,p;i\omega_m,k)}
	{k^2 A^2(i\omega_m,k) +\omega_m^2C^2(i\omega_m,k)+B^2(i\omega_m,k)}
 \,\, , \,\,\, \label{eq:B} \\
 A(i\omega_n,p) &= 1+\frac{2 \pi T}{p^2}\sum_{m = - N}^{N} \int_{0}^{\Lambda} dk
 \frac{k^2 A(i\omega_m,k)E_2(i\omega_n,p;i\omega_m,k)}
	{k^2A^2(i\omega_m,k) +\omega_m^2C^2(i\omega_m,k)+B^2(i\omega_m,k)}
 \,\,,\,\,\,                          \label{eq:A} \\
 C(i\omega_n,p) &= 1+\frac{2 \pi T}{i\omega_n}\sum_{m = - N}^{N}
 \int_{0}^{\Lambda}  dk 
 \frac{k^2 C(i\omega_m,k)E_3(i\omega_n,p;i\omega_m,k)}
	{k^2A^2(i\omega_m,k) +\omega_m^2C^2(i\omega_m,k)+B^2(i\omega_m,k)}
\,\,,\,\,\, \label{eq:C}
\end{align}
where the kernels $E_1$, $E_2$ and $E_3$ are expressed as
\begin{align}
E_1(i\omega_n,p;i\omega_m,k) &=
    - \frac{g^2}{4 \pi^3 p k}
      \ln
\frac{(p-k)^2+(\omega_n-\omega_m)^2}{(p+k)^2+(\omega_n-\omega_m)^2} \,\,, \\
E_2(i\omega_n,p;i\omega_m,k) &= - \frac{g^2}{(2\pi)^3} 
                        \bigg\{ 
			2 +\frac{p^2+k^2+(\omega_n-\omega_m)^2}{2 p k}
 \ln \frac{(p-k)^2
			+(\omega_n-\omega_m)^2}{(p+k)^2+(\omega_n-\omega_m)^2} \bigg\}
			\,\,,\\
E_3(i\omega_n,p;i\omega_m,k) &= 
      -\frac{g^2}{(2\pi)^3}\frac{i\omega_m}{p k}     
  \ln \frac{(p-k)^2+(\omega_n-\omega_m)^2}{(p+k)^2+(\omega_n-\omega_m)^2}\,\,.
\end{align}
Because the integrals in Eqs.~(\ref{eq:B})--(\ref{eq:C}) are divergent, 
we introduce the three-dimensional ultraviolet cutoff $\Lambda$. 
We also truncate the infinite sum of the Matsubara frequency
at a finite number $N$.
However, $N$ can be taken to be sufficiently large so that the 
following results do not depend on it.

We obtain the functions 
$A(\omega,p)$, $B(\omega,p)$ and $C(\omega,p)$ defined at 
discrete points on the imaginary time axis, $\omega=i\omega_n$, 
in the complex $\omega$-plane. 
In the present work we are interested in the fermion spectrum in the chiral 
symmetric phase. 
To specify the chiral symmetric phase, we first determine
the critical temperature, $T_c$, of the chiral phase transition 
from these functions and the effective potential. 
(See Appendix \ref{app:Tc} for details.)
Then, we investigate the fermion spectrum in the chiral symmetric phase, 
in which $B=0$.

Since the fermion spectrum is defined on the real time axis, 
we need to perform an analytic continuation for the solution of 
Eq.~(\ref{eq:ladderSDE}) to the fermion propagator on the real time axis.
Following a method developed in Ref.~\citen{ac},
we perform an analytic continuation of the fermion propagator in the
imaginary time formalism. 
An essential point of this method is the fact that if
the summation of the Matsubara frequency in the loop diagram can be done
analytically, analytic continuation to the retarded function becomes trivial,
i.e. $i\omega_n\to p_0+i\epsilon$.
With this goal, we first perform the summation of the Matsubara
frequency formally, which can be done by expressing the fermion 
and the gauge boson propagators as spectral representations.
Then, the analytic continuation can be done easily, and the remaining
integrals can be carried out numerically.
(For details, see Appendix \ref{app:ac}.)

The equation for the retarded fermion propagator obtained in this way
is the integral equation
\begin{align}
\lefteqn{i S_{R}^{-1}(p_0,\vec{p})-i S_{R\,\,{\rm free}}^{-1}(p_0,\vec{p}) }
\notag \\
&= - g^2 T\sum_{m=-\infty}^{\infty}
\int\frac{d^3\vec{k}}{(2\pi)^3}\int_{-\infty}^{\infty} dz \gamma_{\mu}
\left[\frac{\mathcal{S}(i\omega_m,\vec{k})}{p_0-z-i\omega_m}\right] \gamma_{\nu}
\rho_{B}^{\mu\nu}(z,\vec{p}-\vec{k})
\notag \\
&\hspace{1em} + g^2 \int\frac{d^3\vec{k}}{(2\pi)^3}
\int_{-\infty}^{\infty}dz \gamma_{\mu}
i S_R(p_0-z,\vec{k})\gamma_{\nu} \rho_{B}^{\mu\nu}(z,\vec{p}-\vec{k})\frac{1}{2}
\left[{\rm tanh}\frac{p_0-z}{2T}+{\rm
coth}\frac{z}{2T}\right],
\label{eq:acSDE}
\end{align}
where the subscript $R$ indicates the retarded functions,
$\rho_B^{\mu\nu}$ is the spectral function for the gauge boson, and
$\mathcal{S}(i\omega_m,\vec{k})$ is the solution of Eq.~(\ref{eq:ladderSDE}). 
(The definition of the retarded Green function and the spectral function
are summarized in Appendix \ref{app:note}.)
This method is known to be a reliable method for the analytic continuation in
condensed matter physics and is suitable for the SDE, 
because Eq.~(\ref{eq:ladderSDE}) is also an integral equation that is similar
to Eq.~(\ref{eq:acSDE}).

It is noted here that Eq.~(\ref{eq:acSDE}) is valid for any gauge fixing.
The gauge dependence is in the form of the spectral function, 
$\rho_B^{\mu\nu}$, only. 
In this paper, we adopt the Feynman gauge, in which $\rho_B^{\mu\nu}$ is
expressed by 
\begin{align}
\rho_{B}^{\mu\nu}(p_0,p)
&\equiv 
\frac{1}{\pi}{\rm{Im}}
\left[i D_{R\,\,\rm{free} }^{\mu\nu}(p_0,\vec{p})\right]
\notag\\
&=-g^{\mu\nu}\epsilon(p_0)\delta(p_0^2-p^2)\,\,,
\end{align}
where $\epsilon(p_0)$ is the sign function,
$\epsilon(p_0)=\pm 1$ for $p_0 \gtrless 0$.
For the Landau gauge, on the other hand, the evaluation of the 
double-pole terms in the propagator is rather tedious. 
This is one reason that we adopt the Feynman gauge in this study.

In the chiral symmetric phase, the retarded fermion propagator is written
\begin{align}
-i S_R(p_0,\vec{p})
    &=\frac{\frac{1}{2}(\gamma_0-\vec{\gamma}\cdot\hat{\vec{p}})}{D_+(p_0,p)} 
   +\frac{\frac{1}{2}(\gamma_0+\vec{\gamma}\cdot\hat{\vec{p}})}{D_-(p_0,p)} ,
\end{align}
where $\hat{\vec{p}}=\vec{p}/p$, and $D_{\pm}(p_0,p)$ is defined as
\begin{align}
D_{\pm}(p_0,p) \equiv p_0 C_{R}(p_0,p) \mp p A_{R}(p_0,p)\,\,.
\end{align}
Here, $A_{R}(p_0,p)$ and $C_{R}(p_0,p)$ are the retarded functions 
obtained from $A(i\omega_n,p)$ and $C(i\omega_n,p)$ by analytic continuation, 
respectively.

Performing the angular integrals in Eq.~(\ref{eq:acSDE})
and the projection for the Dirac index,
we obtain the following self-consistent coupled equations for
$A_R(p_0,p)$ and $C_R(p_0,p)$:
\vspace{5em}
\begin{align}
A_R(p_0,p) 
&=  1+\frac{2 \pi T}{p^2}
    \sum_{m = - \infty}^{\infty} \int_{0}^{\Lambda} d k
    \frac{k^2 A(i\omega_m,p) E_2(p_0,p;i\omega_m,k)}
	 {k^2 A^2(i\omega_m,k)+\omega_m^2C^2(i\omega_m,k)} 
 \notag \\
&\hspace{1em} +  
  \frac{g^2}{16 \pi^2 p^3 }\int_{-\infty}^{\infty} d k_0 
    \int_{|p-k_0+p_0|}^{|p+k_0-p_0|} d k 
\notag \\
& \hspace{1em} \times
   \frac{{k} ({p}^2+{k}^2-(p_0-k_0)^2) A_R(p_0,p)X(p_0,k_0;T)}
	{k_0^2 C_R^2(k_0,k)-{k}^2 A_R^2(k_0,k) }  
	 \,\,, \label{eq:acA2}\\
C_R(p_0,p) 
&=  1+\frac{2 \pi T}{p_0}\sum_{m = - \infty}^{\infty} 
	 \int_{0}^{\Lambda} d k
         \frac{k^2 C(i\omega_m,p)
	   E_3(p_0,p;i\omega_m,k) }
	      {k^2 A^2(i\omega_m,k)
		+\omega_m^2C^2(i\omega_m,k)}  
\notag \\
& \hspace{1em}
+ \frac{g^2}{8 \pi^2 p p_0} \int_{-\infty}^{\infty} d k_0 
         \int_{|p-k_0+p_0|}^{|p+k_0-p_0|} d k
	 \frac{{k} k_0 C_R(p_0,p) X(p_0,k_0;T)}
	      {k_0^2 C_R^2(k_0,k)-{k}^2 A_R^2(k_0,k) } 
	      \,\,, \label{eq:acC2}
\end{align}
where $X(p_0,k_0;T)$ is expressed as
\begin{align}
X(p_0,k_0;T) = \tanh\frac{k_0}{2T}+\coth\frac{p_0-k_0}{2T}\,\, .
\end{align}
In the above expressions, we have replaced the integral variable 
$z$ by $k_0=p_0-z$. 
We should note that $A(i\omega_n,p)$ and $C(i\omega_n,p)$  
on the right-hand sides of Eqs.~(\ref{eq:acA2}) and (\ref{eq:acC2}) are
the solutions of the SDE
(\ref{eq:ladderSDE}) obtained in the imaginary time formalism. 

\section{Spectral function}
\label{sec:SDE-sp}
\subsection{Coupling dependence}
In this subsection, we study the fermion spectrum over a wide range of
value of the gauge coupling, focusing on the peak structures.
Because the temperature $T$ is the only infrared scale in the chiral symmetric
phase, dimensional quantities below are scaled by $T$. 
In the following, we fix the cutoff $\Lambda$ as $T/\Lambda=0.3$. 
We have confirmed that the present results depend very little on the choice of 
$\Lambda$. 
(See the end of this section.)

The fermion spectral function is given by 
\begin{align}
  \rho_{\pm}(p_0,p) = -\frac{1}{\pi}{\rm Im}
  \frac{1}{D_{\pm}(p_0,p)} \,\,,
\label{eq:rho-define}
\end{align}
where $\rho_+ ~(\rho_-)$ denotes the fermion (anti-fermion) sector.
The form of $\rho_+ ~(\rho_-)$ can have several peaks, reflecting collective
excitations, e.g. the normal quasi-fermion (quasi-antifermion) and the
anti-plasmino (plasmino) in the HTL approximation. 
We consider only $\rho_+$ in the following, unless otherwise stated, 
noting the relation $\rho_{-}(p_0,p)=\rho_{+}(-p_0,p)$.

\begin{figure}[t]
\begin{center}
	\includegraphics[keepaspectratio,height=6cm]{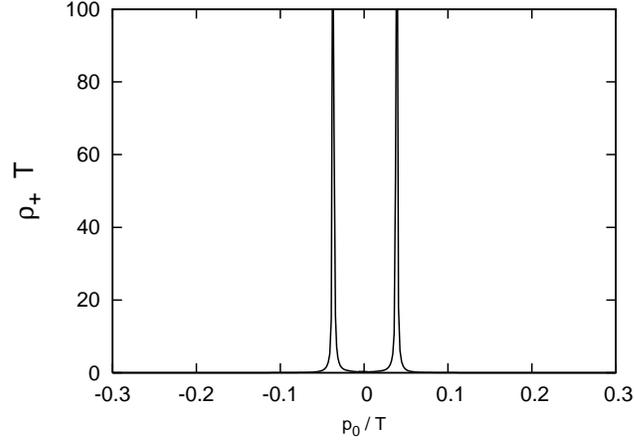}
	\caption{ The spectral function $\rho_+$ at rest ($p=0$) for $g=0.115$. 
	    The sharp peak in the $p_0/T>0 ~(<0)$ region corresponds to
	    the normal quasi-fermion (the anti-plasmino).
	}
	\label{fig:sp-g01}
\end{center}
\end{figure}
\begin{figure}[ht]
\begin{center}
       \includegraphics[keepaspectratio,height=6cm]{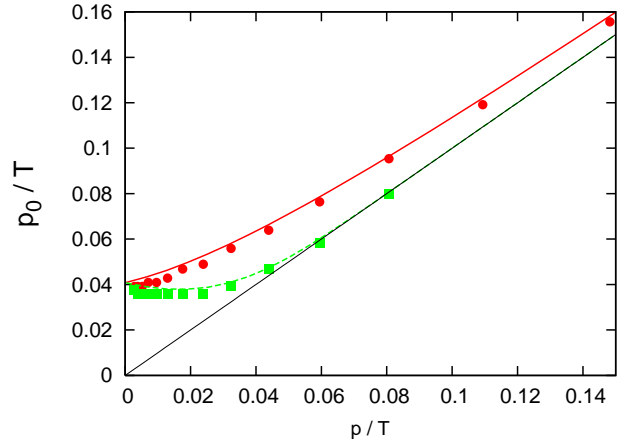}
       \caption{Peak positions of the spectral functions 
	 for the normal quasi-fermion 
      (circles) and the plasmino (squares) for $g=0.115$. 
      These are determined from the peak positions of the spectral function.
	   The solid and the dashed curves represent the dispersion relations
	 for the quasi-fermion and the plasmino in the HTL approximation,
	 respectively. 
	   The thin solid line {corresponding to} $p_0=p$ is also shown
       for convenience.}
       \label{fig:dis-g01}
\end{center}
\end{figure}

We first investigate the fermion spectrum in the weak coupling region.
The spectral function $\rho_+$ for $p=0$ and $g=0.115$ 
is shown in Fig.~\ref{fig:sp-g01}. 
We see that $\rho_{+}$ has two sharp peaks, 
one in the positive energy region, corresponding to the normal
quasi-fermion, and the other in the negative energy region, 
corresponding to the anti-plasmino. 
The peak positions of $\rho_+$ and $\rho_-$ for positive energy 
and finite momentum are plotted in Fig.~\ref{fig:dis-g01},
which displays approximate dispersion relations of these excitations.
We see that the dispersion relations obtained from the SDE
are in good agreement with those obtained with the HTL approximation, 
which is valid in the weak coupling region. 
The narrow widths of the peaks imply that the spectra from the SDE are quite
similar to those in the HTL approximation in the weak coupling region.
This result is natural, because the SDE here contains the fermion self-energy
in the HTL approximation, which is dominant in the weak coupling region 
in our approach, as expected.

\begin{figure}[t]
  \hspace*{-1cm}
  \includegraphics[keepaspectratio,height=4.5cm]{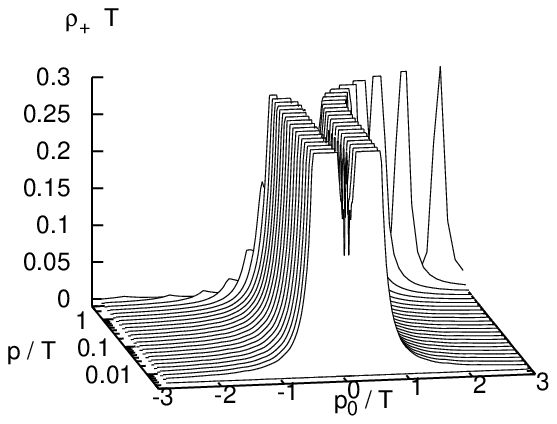}\hspace{-1.7cm}
  \includegraphics[keepaspectratio,height=4.5cm]{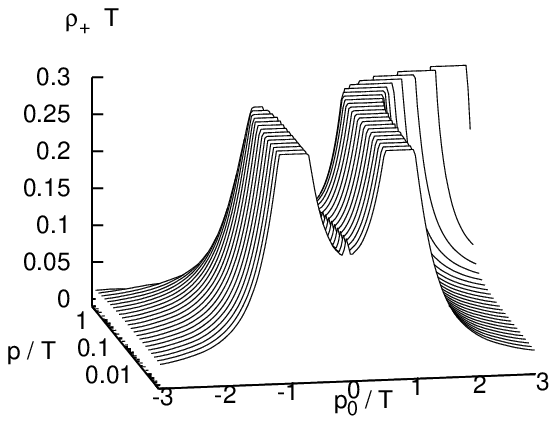}\hspace{-1.7cm}
  \includegraphics[keepaspectratio,height=4.5cm]{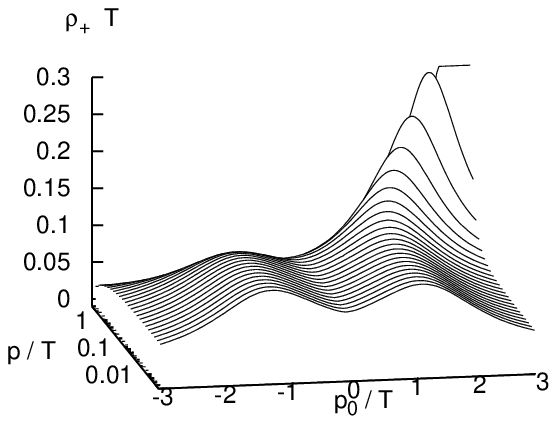}
      \caption{
	The spectral functions $\rho_+$ for  
	$g=1.15$ (left), $g=3$ (middle), and $g=6.9$ (right).
	The peak for $p_0>0 ~(<0)$ corresponds to the normal 
	  quasi-fermion (the anti-plasmino).
      The figure is clipped at $\rho_+ T = 0.3$.}
      \label{fig:sp-g-depnd}
\end{figure}

\begin{figure}[t]
\begin{center}
	\includegraphics[keepaspectratio,height=6cm]{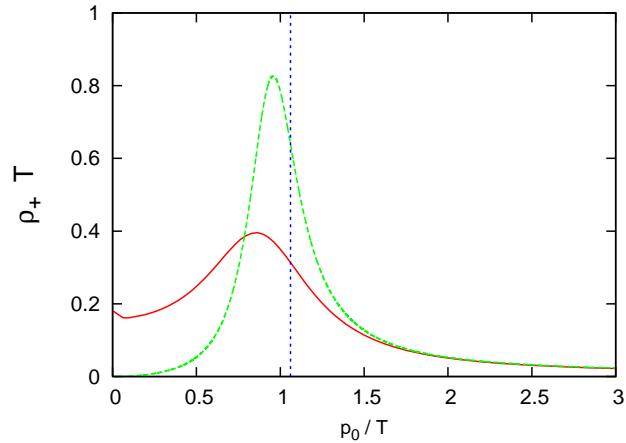}
	\caption{The spectral functions $\rho_+$ at rest ($p=0$) for $g=3$. 
      The solid curve represents the spectral function obtained from the SDE,
      the dashed curve that from the one-loop self-energy, 
      and the dotted line that from the HTL approximation.}
      \label{fig:sp-1}
\end{center}
\end{figure}

\begin{figure}[htbp]
\begin{center}
	\includegraphics[keepaspectratio,height=6cm]{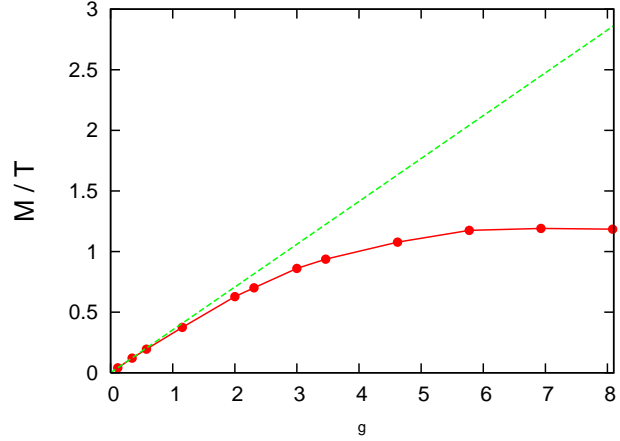}
      \caption{Coupling dependence of the thermal mass. 
      The solid curve represents the thermal mass obtained from the SDE,
      and the dashed line that from the HTL approximation.}
      \label{fig:t03216-Mth-g}
\end{center}
\end{figure}

\begin{figure}[htbp]
\begin{center}
    \includegraphics[keepaspectratio,height=6cm]{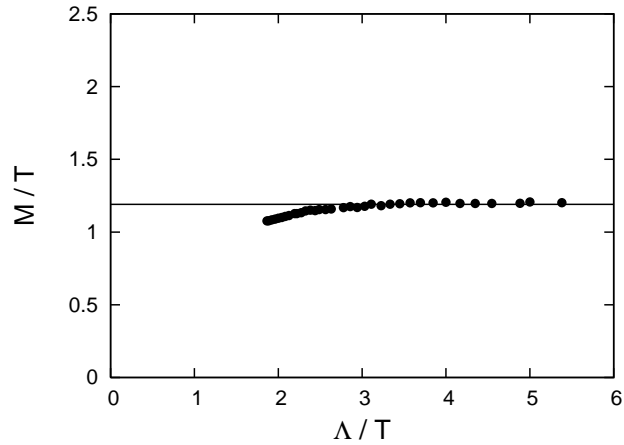}
    \caption{
      Cutoff dependence of the ratio $M/T$ for $g=6.9$. 
      The solid line represents $M/T=1.19$. }
    \label{fig:g6-M-t-detail}
\end{center}
\end{figure}

For the case of strong coupling, we plot the spectral functions $\rho_+$
for $g=1.15, 3$ and $ 6.9$ in Fig.~\ref{fig:sp-g-depnd}.
All the spectral functions appearing in Fig.~\ref{fig:sp-g-depnd}
have two peaks but with broad widths. 
This shows that the normal quasi-fermion and the plasmino appear even in the
strong coupling region. 
We see that as the coupling becomes larger, the widths become broader. 
In the case of extremely strong coupling, e.g. $g=6.9$, the peaks are so broad 
at low momenta that the quasi-particle picture is no longer valid. 
For these three couplings, we have confirmed that as the momentum
becomes higher, the peak of the plasmino becomes smaller, 
and the peak of the normal quasi-fermion becomes larger and sharper. 
This behavior is due to the fact that thermal effects become smaller as
the momentum increases,
and thus the spectrum approaches that for the free fermion.
This momentum dependence of the heights of these peaks
is similar to that of the residues of the peaks in the HTL 
approximation.

In the left and middle panels of Fig.~\ref{fig:sp-g-depnd},
there is another small peak at the origin in addition to the two obvious peaks. 
We comment on this peak in \S \ref{sec:3-peak}.

In Fig.~\ref{fig:sp-1},
we plot the spectral function $\rho_+$ for $g=3$ and $p=0$ in the
positive energy region and compare it with that obtained from the one-loop
self-energy\footnote{Here, the one-loop self-energy includes the
  $T$-independent part  
  evaluated with the cutoff $\Lambda$. In Ref.~\citen{Peshier:1998dy}, 
  on the basis of the one-loop self-energy without the $T$-independent part, 
  it is shown that the thermal mass is slightly larger than that in the HTL 
  approximation, with the difference being less than $5\%$.} 
and that obtained in the HTL approximation. 
This figure shows that the peak obtained at one-loop order
is already broad in the strong coupling region.
It also shows that the peak from the SDE
becomes even broader owing to non-perturbative effects. 
Because the SDE contains effects of multiple scatterings with gauge bosons 
through the self-consistency condition,
the broadening of the peaks could be understood from the fact that the
probability of gauge boson emission and absorption from a fermion increases
rapidly with the coupling.

Now, we interpret the peak positions for $p=0$ as 
the thermal masses of the excitations irrespectively of the widths.
In Fig.~\ref{fig:t03216-Mth-g} we plot the coupling dependence of the
thermal mass together with that obtained in the HTL approximation. 
We note that the thermal masses of the normal quasi-fermion and the plasmino 
are always the same in the chiral limit at zero density\cite{Weldon:1999th}.  
We see that in the weak coupling region, 
the thermal mass from the SDE almost coincides with that in the HTL
approximation, and both of them are proportional to the coupling, $g$.
In the strong coupling region, by contrast, 
the thermal mass from the SDE becomes saturated at $M \sim T$ and is almost
independent of the $g$.

Before ending this subsection, we study the dependence of the resultant
fermion spectrum on the choice of the ultraviolet cutoff $\Lambda$.
In Fig.~\ref{fig:g6-M-t-detail} we show the $\Lambda$-dependence of 
$M/T$ for $g=6.9$.
We see that the ratio is almost constant, $M/T \sim 1.2$, 
for $\Lambda/T \gtrsim 2.5$.
This implies that the thermal mass is determined by only $T$ which 
gives an infrared scale of the system.
%
%
Furthermore, we have confirmed that the fermion spectral functions shown in
Fig.~\ref{fig:sp-g-depnd} have little dependence on the cutoff 
for $p_0 \sim M$. 
The shapes of the spectral function in the large $p_0$ region
are slightly modified when we use smaller cutoff, 
while we see the tendency for the peak of the normal quasi-fermion 
becomes larger and sharper as the momentum becomes higher. 
Thus, we conclude that the thermal mass $M$, as well as the tendency of the
shape, is not affected by cutoff artifacts for $\Lambda/T \gtrsim 2.5$.
For $\Lambda/T \lesssim 2.5$, however, deviation from a constant value 
is seen. There cutoff effects might start to contribute.

\subsection{Comment on the three-peak structure}
\label{sec:3-peak}
\begin{figure}[t]
\begin{center}
	\includegraphics[keepaspectratio,height=6cm]{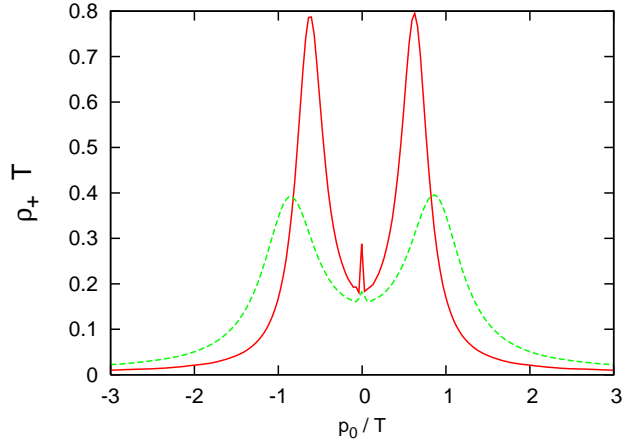}
      \caption{The spectral functions $\rho_+$ at rest ($p=0$) for 
	$g=3$ (dashed curve) and $g=2$ (solid curve).}
      \label{fig:sp-g2-g3-3peak}
\end{center}
\end{figure}
Finally, we give a comment on the peak around the origin
that appears in the spectral function for $g=3$, as seen in the
previous section.
This peak is more pronounced for weaker coupling, $g=2$, as shown in
Fig.~\ref{fig:sp-g2-g3-3peak}.\footnote{In Fig.~\ref{fig:sp-g01}, the spectral
        function near  
	$p_0/T=0$ is not plotted because of the infrared cutoff introduced 
	in the numerical analysis. 
	A sharp peak with a width narrower than the infrared cutoff should
	exist. In the present analysis, we cannot confirm its existence 
	due to numerical error.}
For stronger coupling, such a peak is not seen.
One possible reason for this behavior is that large widths of all the peaks
due to multiple scatterings with gauge bosons may conceal the peak structure
around the origin.

It has been shown in Yukawa models a similar peak appears in the
fermion spectral function owing to the coupling with a massive boson
at finite $T$\cite{Kitazawa:2006zi,Mitsutani:2007gf}. 
In this case, the mass of the boson is essential for the formation of
the three-peak structure in the fermion spectral function.
In our case, however, the fermion couples to a massless gauge boson,
and thus another mechanism makes it possible to form the three-peak
structure.
Because the SDE is a self-consistent equation for the full fermion 
propagator, the internal fermion in the self-energy has a thermal mass.
Therefore, the system is similar to that consisting of a fermion with a 
thermal mass coupled to a massless gauge boson. 
Here, as an example of such a system,
we consider an interaction of a fermion with a thermal mass obtained in the
HTL approximation and a massless gauge boson at one loop.
We summarize the details of this analysis in Appendix \ref{app:HTL-1loop}. 
Figures \ref{fig:ImSigma0} and \ref{fig:sp-HTL} in
Appendix \ref{app:HTL-1loop} plot the imaginary part of the fermion
self-energy and the spectral function at rest in this model.
We see that the imaginary part of the self-energy possesses two peaks, 
which lead to the three-peak structure in the spectral function, 
as shown in Refs.~\citen{Kitazawa:2006zi} and \citen{Mitsutani:2007gf}. 
Therefore, the existence of the three-peak structure of the spectral function 
from the SDE can be understood from this result.

\section{Summary and discussion}
\label{sec:sum}
We have investigated the fermion spectrum in the chiral symmetric phase 
employing the Schwinger-Dyson equation (SDE) for the fermion propagator with
the ladder approximation and fixed gauge coupling, 
in which the free gauge boson propagator in the Feynman gauge is used. 
Because we employed the imaginary time formalism to construct the SDE, 
analytic continuation to the real time axis is needed to obtain the fermion 
spectrum. 
This was done by solving an integral equation for the retarded fermion 
propagator, following a method developed in Ref.~\citen{ac}. 
This method is known as a reliable way to carry out numerical analytic
continuations in condensed matter physics, and we have confirmed in this study 
that it is also effective even in a relativistic system.

We have investigated the fermion spectrum in the chiral symmetric phase 
over a wide range of values of the gauge coupling using the method of analytic
continuation mentioned above. 
We found that in the weak coupling region ($g\ll 1$), 
the fermion spectrum is quite similar to that obtained in the hard thermal
loop (HTL) approximation: 
There exist a normal quasi-fermion and a plasmino whose thermal masses  
are almost the same as those found in the HTL approximation, 
$M \sim M_{\rm HTL}= g T/\sqrt{\mathstrut 8}$, and their widths are very
narrow. (These widths vanish in the HTL approximation.) 
This result is natural, because the SDE here contains the HTL of the fermion
self-energy, and thus the HTL is dominant in the weak coupling region 
in our approach, as expected.

In the strong coupling region $(g>1)$, we find that there exist 
a normal quasi-fermion and a plasmino, as in the weak coupling region. 
However, their spectra deviate from those in the HTL approximation:
Both the excitation modes have smaller thermal masses and much broader widths
in the low-momentum region. 
The thermal mass $M$ saturates at some value of the gauge coupling,
and satisfies $M\sim T$ {\it independently of the value of the coupling}. 
Although we do not have any plausible explanation for the
saturation of the thermal mass at present,
the saturated value $M\sim T$ is consistent with the results of 
other model analyses \cite{Mannarelli:2005pz,Schaefer:1998wd,%
Kitazawa:2005mp,Karsch:2007wc,Kitazawa:2006zi}. 
In Ref.~\citen{Mannarelli:2005pz}, 
the broad width is described in terms of the imaginary part of the self-energy, 
while in our study, it is described in terms of the spectral function. 
Furthermore, we have shown that as the gauge coupling becomes larger, 
the widths rapidly become broader. 
This indicates that the broadening of the widths can be understood from the
fact that the probability of gauge boson emission and absorption from a
fermion increases rapidly with the coupling.

Our result shows that as the momentum $p$ becomes higher, 
the peak of the plasmino becomes smaller and the peak of the normal 
quasi-fermion becomes higher and sharper. 
Thus the spectrum approaches the free-fermion one. 
This behavior is similar to that in the weak coupling region and is natural
because thermal effects are weaker in general for $p > T$. 
The free-fermion spectrum in the high momentum region is consistent with 
the quark recombination or coalescence picture,\cite{Fries:2003vb} in which 
the quasi-quark picture is valid.

One reason we have adopted the Feynman gauge is for the simplicity of the
numerical analytic continuation mentioned above.
We expect that the qualitative features of the fermion spectrum discussed in
this paper do not change even if we adopt other gauge fixings.

We now discuss what our result can provide regarding the
quark spectrum in the real-life QCD at high temperature.
In the very high temperature region of QCD, 
the HTL approximation is valid for studying the quark spectrum.
In the HTL approximation, the fixed coupling, which is the
running coupling at an energy scale of order $T$, 
say $g(T)$, is used, because the quarks and gluons with energy $T$ give
the dominant contribution.
In studying the quark spectrum in the lower $T$ region,
the diagrams other than the HTL start to give non-negligible corrections, 
since the (fixed) coupling $g(T)$ is larger.
For this reason, we can state that, in our approach, 
we include a part of the corrections
from all the ladder diagrams by solving the Schwinger-Dyson
equation.
Actually, our result reproduces that of the HTL approximation
in the weak coupling region, i.e. in the very high temperature
region where $g(T)$ is small.
In the lower temperature region, where $g(T)$ is large, 
our result in the strong coupling region suggests
that the corrections from the ladder diagrams have 
the following effects on 
the qualitative structure of the quark spectrum:
(1) The thermal masses of quark and plasmino
are not proportional to the gauge coupling $g(T)$, becoming $M \sim T$, and 
(2) their widths are very broad, due to multiple scatterings with gluons.

In this study, we have used the free gauge boson propagator in the SDE. 
In medium, however, the gauge boson spectrum may also change 
significantly, as in the HTL approximation.
One possible extension to include such an effect is to use the 
gauge boson propagator 
with Debye screening and a dynamical screening that has the same
form as that in the HTL approximation.
Such a study is left as a future project. 

\section*{Acknowledgements} 
This work is supported in part by the 21st Century COE Program at
Nagoya University, the Daiko Foundation \#9099 (M.~H.), and a JSPS
Grant-in-Aid for Scientific Research, \#18740140 (Y.~N.). 

\appendix
\section{Notation}
\label{app:note}
Here, we briefly give definitions of some of the quantities used in this 
paper.

The Matsubara Green function is defined by 
\begin{align}
\mathcal{S}(i\omega_n,\vec{p})
            \equiv {\rm F.T.}\langle T_{\tau}\psi(\tau,\vec{r})\overline{\psi}
             (\tau^{\prime},\vec{r^{\prime}})\rangle_{T} \,\,,
\end{align}
where $\langle\cdots\rangle_T$ stands for the thermal average
with temperature $T$. 
The retarded Green function for a fermion $S_R$ and the spectral function
for a fermion are defined by
\begin{align}
S_{R}(t,\vec{r};t^{\prime},r^{\prime})
           \equiv\theta(t-t^{\prime})\langle \{\psi(t,\vec{r}),\overline{\psi}
             (t^{\prime},\vec{r^{\prime}}) \} \rangle_{T}\,\,, \\
\rho(p_0,p)
           =\frac{1}{2 \pi} {\rm F.T.} \langle \{\psi(t,\vec{r}),\overline{\psi}
             (t^{\prime},\vec{r^{\prime}}) \} \rangle_T \,\,.
\end{align}
The retarded Green function and the Matsubara Green function are related to 
the spectral function by the spectral representations as
\begin{align}
\mathcal{S}(i\omega_n,\vec{p})&=-\int_{-\infty}^{\infty}d
p_0^{\prime}\frac{\rho(p_0^{\prime},\vec{p})}{i\omega_n-p_0^{\prime}} \,\,,\\
iS_{R}(p_0,\vec{p})&=-\int_{-\infty}^{\infty}d
p_0^{\prime}\frac{\rho(p_0^{\prime},\vec{p})}{p_0-p_0^{\prime}+i\epsilon} \,\,.\
\end{align}
These functions for gauge bosons can be written in the same forms.

\section{Phase Structure}
\label{app:Tc}
\begin{figure}[t]
	\includegraphics[keepaspectratio,height=5cm]{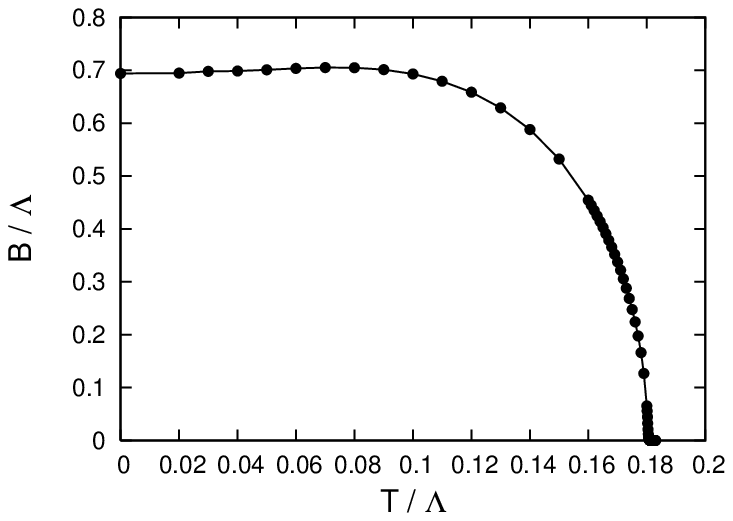}\hspace{-3mm}
	\includegraphics[keepaspectratio,height=5cm]{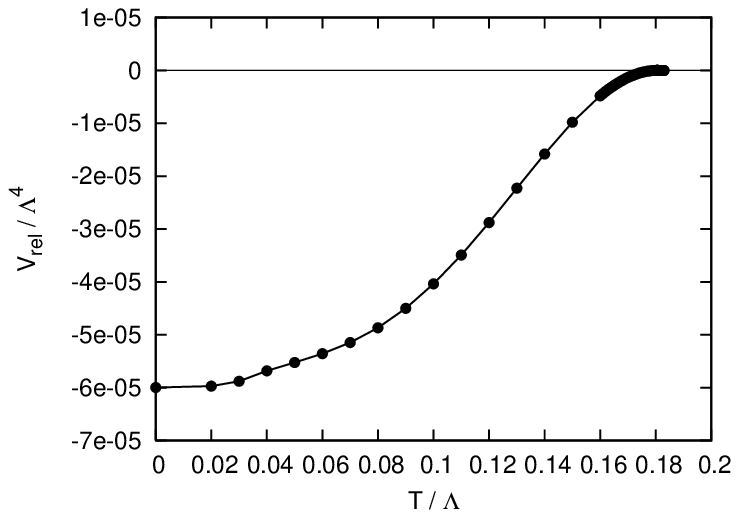}
      \caption{Temperature dependence of the function
      $B(i\omega_n=0,p=0)/\Lambda$ (left) and the effective potential
      $V_{\rm rel}(T)/\Lambda^4$ (right) for $g=6.9$.}
      \label{fig:B-t}
\end{figure}
In this appendix, we compute the critical temperature $T_c$ to specify the
chiral symmetric phase. 
We introduce the effective potential for the fermion propagator
to determine the phase structure and $T_c$,\cite{Cornwall:1974vz}
\begin{align}
V(\mathcal{S})&=  
  T\sum_{n=-\infty}^{\infty}  \int\frac{d^3\vec{p}}{(2\pi^3)}
  \{\ln \det [\mathcal{S}(i\omega_n,\vec{p})]+{\rm tr}[\Sfl{p}
    \mathcal{S}(i\omega_n,\vec{p})]\} 
\notag\\
&+ \frac{1}{2} g^2T^2\sum_{n,m=-\infty}^{\infty}
   \int\frac{d^3\vec{p}}{(2\pi)^3}\int\frac{d^3\vec{k}}{(2\pi)^3} 
\notag \\
&\hspace{2em} \times
   {\rm tr}\left[   
     \mathcal{S}(i\omega_n,\vec{p})\gamma_{\mu}
     \mathcal{S}(i\omega_m,\vec{k})\gamma_{\nu}\right] 
 \mathcal{D}^{\mu\nu}(i\omega_n-i\omega_m,\vec{p}-\vec{k})  \,\,,
 \label{eq:v}
\end{align}
where ${\rm ln}$, $\det$ and ${\rm tr}$ are taken in the spinor spaces. 
The SDE is obtained as the stationary condition for
the effective potential (\ref{eq:v}). 
Note that the value of the potential is meaningful only at the
stationary point; i.e. the potential can be evaluated only when the 
solution of the SDE is substituted.
There always exists a solution of the SDE with $B=0$. 
In the low temperature region, we also have a solution with $B\neq 0$, 
which corresponds to dynamical chiral symmetry breaking.
Let $\mathcal{S}_{W}$ denote the solution with $B=0$
and $\mathcal{S}_{NG}$ the solution with $B\neq 0$. 
They are parameterized as
\begin{align}
  [\mathcal{S}_W(i\omega_n,\vec{p})]^{-1} &\equiv 
  A_W(i\omega_n,p)\vec{p}\cdot\vec{\gamma}-C_W(i\omega_n,p)i\omega_n\gamma_0
  \label{eq:Sw} \,\,,\\
  [\mathcal{S}_{NG}(i\omega_n,\vec{p})]^{-1} &\equiv 
  B_{NG}(i\omega_n,p)
  +A_{NG}(i\omega_n,p)\vec{p}\cdot\vec{\gamma}
  -C_{NG}(i\omega_n,p)i\omega_n\gamma_0 
  \label{eq:Sng}\,\,.
\end{align}
To determine which of the solutions $\mathcal{S}_W$ and $\mathcal{S}_{NG}$
gives the real vacuum, we compare the values of the effective potential 
with each solution by computing the following difference:
\begin{align}
V_{\rm rel}(T) 
&\equiv V(\mathcal{S}_{NG})-V(\mathcal{S}_{W})\notag\\
&= -\frac{1}{\pi^2}T\sum_{m=-\infty}^{\infty} \int_{0}^{\Lambda} dp p^2 
   \Biggl\{\ln\left[\frac{p^2 A_{NG}^2(i\omega_n,p)+\omega_n^2
   C_{NG}^2(i\omega_n,p)+B_{NG}^2(i\omega_n,p)}
     {p^2 A_{W}^2(i\omega_n,p)+\omega_n^2
       C_{W}^2(i\omega_n,p)} \right]    
\notag \\ 
&\  + \frac{p^2 A_{NG}(i\omega_n,p)+\omega_n^2 C_{NG}(i\omega_n,p)}
   {p^2 A_{NG}^2(i\omega_n,p)+\omega_n^2
     C_{NG}^2(i\omega_n,p)+B_{NG}^2(i\omega_n,p)} 
\notag \\
&\  - \frac{p^2 A_{W}(i\omega_n,p)+\omega_n^2 C_{W}(i\omega_n,p)}
  {p^2 A_{W}^2(i\omega_n,p)+\omega_n^2 C_{W}^2(i\omega_n,p)}
  \Biggr\}  \,\,.\,\,
\end{align}
For $V_{\rm rel}(T)<0$, the chiral broken vacuum is realized,
while the chiral symmetric phase becomes the true vacuum for
$V_{\rm rel}(T)\geq 0$.
We determine $T_c$ as the smallest value of $T$ at which  
$V_{\rm rel}(T)$ vanishes. 

\begin{figure}[t]
\begin{center}
	\includegraphics[keepaspectratio,height=6cm]{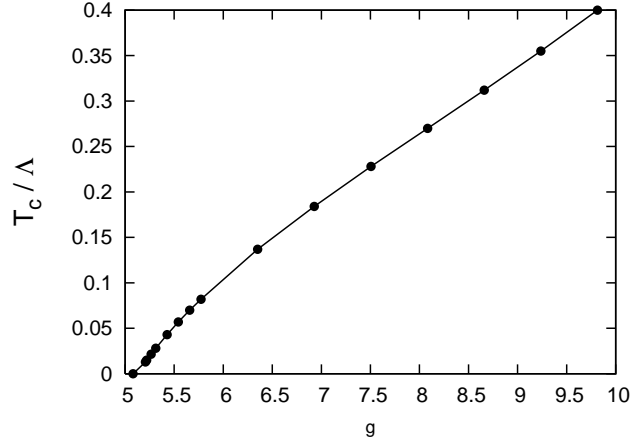}
      \caption{Coupling dependence of the critical temperature
        $T_c/\Lambda$.
	The chiral symmetric (broken) phase is above (below) the line. }
      \label{fig:tc-g}
\end{center}
\end{figure}

We plot the $T$ dependence of the function $B(i\omega_n,p)$
for the smallest Matsubara frequency at $p=0$ 
in the left panel and that of the potential $V_{\rm rel}$ in the right panel
of Fig.~\ref{fig:B-t}.
It is seen that the function $B$ and the potential $V_{\rm rel}$
simultaneously go to zero smoothly with $T$. 
This implies that the chiral phase transition is of second order.
The value of the critical temperature is determined as 
$T_c/\Lambda \simeq 0.18$ for $g=6.9$. 
The value $T_c/\Lambda$ depends on the coupling. 
In Fig.~\ref{fig:tc-g}, we plot $T_c/\Lambda$ for each coupling.
Note that the chiral broken vacuum is not realized even at $T=0$ for 
$g<g_c\simeq 5.1$. 

\section{Analytic Continuation for Solutions \\
of the Schwinger-Dyson Equation}
\label{app:ac}
In this appendix, we explain the details of the method for the analytic
continuation used in the present analysis, following Ref.~\citen{ac}.

Let us start with the spectral representations for the fermion and gauge boson
propagators, 
\begin{align}
\mathcal{S}(i\omega_m,\vec{k}) 
&= -\int_{-\infty}^{\infty}dz
    \frac{\rho_F(z,\vec{k})}{i\omega_m-z}
= -\frac{1}{\pi}\int_{-\infty}^{\infty}dz
	\frac{{\rm Im} i S_R(z,\vec{k})}{i\omega_m-z} \,\, ,\\
\mathcal{D}_{\mu\nu}(i\omega_n-i\omega_m,\vec{p}-\vec{k}) 
&= -\int_{-\infty}^{\infty}dz
\frac{\rho_B^{\mu\nu}(z,\vec{p}-\vec{k})}{(i\omega_n-i\omega_m)-z} \,\, ,
\end{align}
where $\rho_F$ and $\rho_B^{\mu\nu}$ represent the spectral functions of
fermion and the gauge boson, respectively. 
The function $S_R(z,\vec{k})$ is the retarded fermion propagator, which is
analytic in the upper half complex plane of the variable $z$.

Using the above spectral representations, the summation over the Matsubara
frequencies in the SDE (\ref{eq:ladderSDE}) is performed
by replacing the sums over frequencies with the following contour integral:
\begin{equation}
T\sum_{m=-\infty}^{\infty} f(k_0 = i\omega_m) = \oint_c \frac{dk_0}{2\pi
 i}f(k_0)\frac{1}{2}{\rm tanh}\frac{k_0}{2T} \,\,. 
\end{equation}
Then we replace the Matsubara frequency as
$i\omega_n\rightarrow p_0+ i\epsilon$.
Thus we obtain the SDE for the retarded fermion propagator as follows:
\begin{align}
&iS_{R}^{-1}(p_0,\vec{p})-iS_{R\,\,{\rm{free}}}^{-1}(p_0,\vec{p})   
\notag \\
&= - \frac{g^2}{\pi}\int \frac{d^3 k}{(2\pi)^3} 
  \int_{-\infty}^{\infty}dz  \int_{-\infty}^{\infty}dz^{\prime} 
\notag \\
&\ \times  \Biggl[\gamma_{\mu}\frac{{\rm Im}[i S_R(z,\vec{k})]}
  {p_0-z-z^{\prime}+i\epsilon}\gamma_{\nu}
  \rho_B^{\mu\nu}(z^{\prime},\vec{p}-\vec{k})
  \frac{1}{2}\left(\tanh\frac{z}{2T}+\coth
  \frac{z^{\prime}}{2T} \right)  \Biggr] \,\,.
\label{eq:acSDE-nonz}
\end{align}

By using the relations ${\rm Re}[f(z)]={\rm Im}[i f(z)]$ and 
${\rm Im}[f(z)]={\rm Re}[-if(z)]$, Eq.~(\ref{eq:acSDE-nonz}) can be rewritten
as 
\begin{align}
&\lefteqn{iS_{R}^{-1}(p_0,\vec{p})-iS_{R\,\,{\rm{free}}}^{-1}(p_0,\vec{p})}   \nonumber \\
&= -\frac{g^2}{4\pi}
   \int_{-\infty}^{\infty}dz^{\prime}\int\frac{d^3\vec{k}}{(2\pi)^3}
   \rho_B^{\mu\nu}(z^{\prime},\vec{p}-\vec{k})\gamma_{\mu} \notag \\
& \hspace{1em} \times
   \Biggl\{ {\rm Im}\Biggl[\int_{-\infty}^{\infty}dz iS_R(z,\vec{k}) 
     \left(\frac{1}{p_0-z-z^{\prime}+i\epsilon}
     +\frac{1}{p_0-z-z^{\prime}-i\epsilon}\right)  \notag\\  
& \hspace{3em} \times 
     \left({\rm tanh}\frac{z}{2T}                            
     +{\rm coth}\frac{z^{\prime}}{2T} \right) \Biggr]  \notag \\
& \hspace{3em}
   -i{\rm Re}\Biggl[\int_{-\infty}^{\infty}dz iS_R(z,\vec{k})
     \left(\frac{1}{p_0-z-z^{\prime}+i\epsilon}
     -\frac{1}{p_0-z-z^{\prime}-i\epsilon}\right) \notag\\
& \hspace{3em} \times  \left({\rm
       tanh}\frac{z}{2T} +{\rm coth}\frac{z^{\prime}}{2T} \right) \Biggr]
   \Biggr\} \gamma_{\nu} \,\,.
\label{eq:acSDE-sep}
\end{align}
Because the retarded fermion propagator, $S_R$, should be analytic in the
upper half complex $z$ plane, the integrand of the right-hand side can have 
poles only at $z=p_0-z^{\prime}+i\epsilon$ and
$z=i(2n+1)\pi T$ for $n\geq 0$. 
Then, the integration over $z$ can be replaced by the contour integral. 
Then, performing the contour integral over $z$, we obtain
\begin{align}
&\lefteqn{iS_{R}^{-1}(p_0,\vec{p})-iS_{R\,\,{\rm{free}}}^{-1}(p_0,\vec{p})}   
\notag \\
&= - \frac{g^2}{2\pi} 
\int_{-\infty}^{\infty}dz^{\prime}\int\frac{d^3\vec{k}}{(2\pi)^3}
\rho_B^{\mu\nu}(z^{\prime},\vec{p}-\vec{k}) \gamma_{\mu} \notag \\
&\hspace{1em} \times 
\Biggl\{ {\rm Im}\Biggl[2i\pi T \sum_{m=0}^{\infty} i S_R(i\omega_m,\vec{k}) 
  \left(\frac{1}{p_0-i\omega_m-z^{\prime}+i\epsilon}
  +\frac{1}{p_0-i\omega_m-z^{\prime}-i\epsilon}\right)    
  \notag \\
&\hspace{5em}-i\pi i S_R(p_0-z^{\prime},\vec{k})\left({\rm tanh}\frac{p_0-z^{\prime}}{2T}
  +{\rm coth}\frac{z^{\prime}}{2T} \right)
  \Biggr]  \notag \\
&\hspace{3em}
-i{\rm Re}\Biggl[2i\pi T \sum_{m=0}^{\infty} i S_R(i\omega_m,\vec{k})
  \left(\frac{1}{p_0-i\omega_m-z^{\prime}+i\epsilon}
  -\frac{1}{p_0-i\omega_m-z^{\prime}-i\epsilon}\right)
  \notag \\
&\hspace{6em}
  -i\pi i S_R(p_0-z^{\prime},\vec{k})
  \left({\rm tanh}\frac{p_0-z^{\prime}}{2T} +{\rm coth}\frac{z^{\prime}}{2T}
  \right)  \Biggr]
\Biggr\} \gamma_{\nu} \,\,.\\
\nonumber
\end{align}
In the above expression, we can drop the term $i\epsilon$, 
because the Matsubara frequency, $\omega_m$, is always non-zero. 
Using the relations 
${\rm Im} f(z) -i {\rm Re}f(z)=-if(z)$ and ${\rm Im}f(z)+i{\rm Re}f(z)=if(z)^{*}$, 
we obtain
\begin{align}
&iS_{R}^{-1}(p_0,\vec{p})-iS_{R\,\,{\rm{free}}}^{-1}(p_0,\vec{p})
\notag \\
&= - g^2\int_{-\infty}^{\infty}dz^{\prime} \int \frac{d^3\vec{k}}{(2\pi)^3}
\rho_B^{\mu\nu}(z^{\prime},\vec{p}-\vec{k})\gamma_{\mu} 
\Biggl[T\sum_{m=0}^{\infty}
\left(\frac{S(i\omega_m,\vec{k})}{p_0-i\omega_m-z^{\prime}}
+\frac{S^{*}(i\omega_m,\vec{k})}{p_0-i\omega_m-z^{\prime}}\right) \notag\\
&\ 
- i S_R(p_0-z^{\prime},\vec{k})\frac{1}{2}
\left({\rm tanh}\frac{p_0-z^{\prime}}{2T}
+{\rm coth}\frac{z^{\prime}}{2T}\right)\Biggr]\gamma_{\nu} \,\,.
\label{eq:acSDE-c}
\end{align}
Finally, because the Matsubara Green function is real, 
i.e. $S(i\omega_m,\vec{k})=S(i\omega_m,\vec{k})^{*}$, 
we arrive at
\begin{align}
\lefteqn{i S_{R}^{-1}(p_0,\vec{p})-i S_{R\,\,{\rm free}}^{-1}(p_0,\vec{p}) }
\notag \\
&= - g^2\int\frac{d^3\vec{k}}{(2\pi)^3}\int_{-\infty}^{\infty} dz 
   T\sum_{m=-\infty}^{\infty}
   \gamma_{\mu}
   \left[\frac{\mathcal{S}(i\omega_m,\vec{k})}{p_0-z-i\omega_m}\right] 
   \gamma_{\nu}
   \rho_{B}^{\mu\nu}(z,\vec{p}-\vec{k})
\notag \\
&\hspace{1em}+ 
   g^2\int\frac{d^3\vec{k}}{(2\pi)^3}
   \int_{-\infty}^{\infty}dz 
   \gamma_{\mu}
   i S_R(p_0-z,\vec{k})\gamma_{\nu} 
   \rho_{B}^{\mu\nu}(z,\vec{p}-\vec{k})\frac{1}{2}
   \left[{\rm tanh}\frac{p_0-z}{2T}+{\rm coth}\frac{z}{2T}\right] \,\,.
\label{eq:acSDE-app}
\end{align}
With the solution of Eq.~(\ref{eq:ladderSDE}) substituted into 
${\mathcal S}(i\omega_m,\vec{k})$,  
Eq.~(\ref{eq:acSDE-app}) 
becomes a self-consistent equation for the retarded fermion propagator. 
In this way, we can perform the analytic continuation for the solution of the
SDE by solving the integral equation (\ref{eq:acSDE-app}).

Note that the first term in Eq.~(\ref{eq:acSDE-app}) 
is obtained through the naive replacement 
$i\omega_n \rightarrow p_0 + i \epsilon$ in Eq.~(\ref{eq:ladderSDE}), 
which is easily checked by integrating over $z$ in Eq.~(\ref{eq:acSDE-app}). 
The second term appears from the fact that the Matsubara summation must be
done before the analytic continuation. 
We can also verify that Eq.~(\ref{eq:acSDE-app}) returns to the original SDE
(\ref{eq:ladderSDE}) through the replacement $p_0 \rightarrow i\omega_n$. 

\section{Three-Peak Structure in a Simple Model}
\label{app:HTL-1loop}
In this appendix, we show that the three-peak structure of the fermion 
spectrum obtained in the SDE can be understood from the interaction of
a fermion with a thermal mass and a free gauge boson. 
For simplicity, let us consider a one-loop diagram consisting of a fermion 
propagator obtained in the HTL approximation and a massless gauge boson propagator. 
Then, the fermion self-energy is given by
\begin{align}
\Sigma(i\omega_n,\vec{p}) 
=  -  g^2 T\sum\int \frac{d^3 k}{(2 \pi)^3}
   \gamma_{\mu} {\mathcal S}^{\rm HTL}(i\omega_m,\vec{k}) \gamma_{\nu} 
   {\mathcal D}^{\mu\nu}_{\rm free}(i\omega_n-i\omega_m,\vec{p}-\vec{k})\,\,,
\label{eq:HTL-self}
\end{align}
where ${\mathcal S}^{\rm HTL}$ is the fermion propagator obtained in the HTL
approximation and ${\mathcal D}_{\rm free}^{\mu\nu}$ is the free gauge boson
propagator in the Feynman gauge. 
The retarded fermion propagator in the HTL approximation is decomposed as  
\begin{align}
-i S_R^{\rm HTL}(p_0,\vec{p})
=\frac{\frac{1}{2}(\gamma_0-\vec{\gamma}\cdot\hat{\vec{p}})}{D_{+}^{\rm HTL}(p_0,\vec{p})}
+\frac{\frac{1}{2}(\gamma_0+\vec{\gamma}\cdot\hat{\vec{p}})}{D_{-}^{\rm HTL}(p_0,\vec{p})}\,\,,
\end{align}
 where $D_+^{\rm HTL}$ is for the fermion and $D_-^{\rm HTL}$ for the anti-fermion. 
The quantities $D_{\pm}^{\rm HTL}$ are related to the spectral function $\rho_{\pm}^{\rm HTL}$ as in Eq.~(\ref{eq:rho-define}).
The spectral functions $\rho_{\pm}^{\rm HTL}$  can be written as 
\begin{align}
\rho^{\rm HTL}_{\pm}(p_0,p) 
  = Z_{\pm}(p)\delta(p_0-\omega_{\pm}(p))
  + Z_{\mp}(p)\delta(p_0+\omega_{\mp}(p))
  + \rho_{\pm}^c(p_0,p)\,\,,  
\label{eq:spc-HTL}
\end{align}
where $Z_{\pm}(k)=(\omega_{\pm}^2(k)-k^2)/2M_{\rm HTL}^2$ are pole residues,
with $M_{\rm HTL}^2= g^2 T^2 /8$, and $\rho_{\pm}^c$ are the continuum parts,
given by 
\begin{align}
\rho^{c}_{\pm}(p_0,p) 
&= \frac{1}{2 p}M_{\rm HTL}^2(1\mp x)\theta(1-x^2) 
\notag \\
&\times 
  \left[
    (p(1\mp x)\pm\frac{M_{\rm HTL}^2}{2p}
    [(1\mp x)\ln \left|\frac{x+1}{x-1}\right|\pm 2])^2
    + \frac{\pi^2 M_{\rm HTL}^4}{4p^2}(1\mp x)^2 \right]^{-1}
\,\,,
\end{align}
with $x=p_0/p$.
In the following, we limit ourselves to the rest frame ($p=0$). 
Using $\rho_{\pm}^{\rm HTL}$, 
we obtain
\begin{align}
\Sigma_0(i\omega_n,0) 
&= \frac{1}{4}{\rm tr}[\gamma_0 \Sigma(i\omega_n,0)] 
\notag\\
&= - g^2 T \sum \int \frac{d^3 k}{(2\pi)^3}
\int d z  
    \frac{\rho^{\rm HTL}_{+}(z,{k})+\rho^{\rm HTL}_{-}(z,{k})}{i\omega_m-z}
\int d z^{\prime}
\frac{\rho_B(z^{\prime},{k})}{i\omega_n-i\omega_m-z^{\prime}} \,\,,
\end{align}
with $\rho_B(p_0,p)=\epsilon(p_0)\delta(p_0^2-p^2)$.
By carrying out the Matsubara summation and the replacement  
$i\omega_n \rightarrow p_0 + i\epsilon$, $\Sigma_0$ becomes
\begin{align}
\Sigma_0(p_0,0) &=  g^2 \int \frac{d^3 k}{(2 \pi)^3}\int dz \int dz^{\prime} 
       \frac{1-f(z)+n(z^{\prime})}{p_0-z-z^{\prime}+i\epsilon} \notag\\
&\ \times       \left[\rho^{\rm HTL}_{+}(z,{k})
	 +\rho^{\rm HTL}_{-}(z,{k}) \right]
       \rho_B(z^{\prime},{k})\,\,,
\label{eq:Sigma0}
\end{align}
with $f(E)=[\exp(E/T)+1]^{-1}$ and $n(E)=[\exp(E/T)-1]^{-1}$. 
The imaginary part of $\Sigma_0$ can be easily calculated as
\begin{align}
{\rm Im} \Sigma_0 
&= \frac{ g^2}{4\pi} \Biggl[
\frac{k
  Z_+(k)}{|1+\omega_+^{\prime}|}(f(p_0-k)-n(k)-1) \Big|_{p_0-k-\omega_+(k)=0} 
\notag\\
&\ + 
\frac{k
  Z_-(k)}{|1-\omega_-^{\prime}|}(f(p_0-k)-n(k)-1) \Big|_{p_0-k+\omega_-(k)=0}  
\notag \\
&\ + \frac{k
  Z_-(k)}{|1+\omega_-^{\prime}|}(f(p_0-k)-n(k)-1) \Big|_{p_0-k-\omega_-(k)=0} 
\notag\\
&\ + 
\frac{k
  Z_+(k)}{|1-\omega_+^{\prime}|}(f(p_0-k)-n(k)-1) \Big|_{p_0-k+\omega_+(k)=0}  
\notag \\
&\ -\frac{k
  Z_+(k)}{|1-\omega_+^{\prime}|}(f(p_0+k)+n(k)) \Big|_{p_0 +k-\omega_+(k)=0} 
\notag\\
&\ - 
\frac{k
  Z_-(k)}{|1+\omega_-^{\prime}|}(f(p_0+k)+n(k)) \Big|_{p_0 +k+\omega_-(k)=0}  
\notag \\
&\ -\frac{k
  Z_-(k)}{|1-\omega_-^{\prime}|}(f(p_0+k)+n(k)) \Big|_{p_0 +k-\omega_-(k)=0} 
\notag\\
&\ - 
\frac{k
  Z_+(k)}{|1+\omega_+^{\prime}|}(f(p_0+k)+n(k)) \Big|_{p_0+k+\omega_+(k)=0} 
\Biggr] 
\notag \\
&\ + \frac{ g^2}{4\pi} 
\int dk k \Bigl[
  (f(p_0-k)-n(k)-1)(\rho_+^c(p_0-k,k)+\rho_-^c(p_0-k,k))
\notag \\
&\hspace{70pt}-(f(p_0+k)+n(k))(\rho_+^c(p_0+k,k)+\rho_-^c(p_0+k,k))\Bigr] \,\,,
\label{eq:ImSigma0}
\end{align}
with $\omega_{\pm}^{\prime}=d\omega_{\pm}(k)/d k$. 

It is noted that Eq.~(\ref{eq:ImSigma0}) is divergence-free. 
The real part of $\Sigma_0$, by contrast, has an ultraviolet divergence, 
and here we regularize it using the relation 
\begin{align}
{\rm Re} \Sigma_0(p_0,0) 
= -\frac{1}{\pi}{\rm P} \int_{-\Lambda}^{\Lambda}d p_0^{\prime}
                 \frac{{\rm Im}\Sigma_0(p_0^{\prime},0)}{p_0-p_0^{\prime}}\,\,,
\label{eq:ReSigma0}
\end{align}
with the three-momentum cutoff $\Lambda$.
We choose $\Lambda$ to satisfy $T/\Lambda =0.3$. 
Here, ${\rm P}$ represents the principal-value operator. 
By using $\Sigma_0$, the spectral function at rest is obtained as 
\begin{equation}
\rho_{\pm}(p_0,0) =  -\frac{1}{\pi}{\rm Im}\frac{1}{p_0-\Sigma_0(p_0,0)}\,\,.
\label{eq:HTL2spc}
\end{equation}

\begin{figure}[t]
\begin{center}
 	\includegraphics[keepaspectratio,height=6cm]{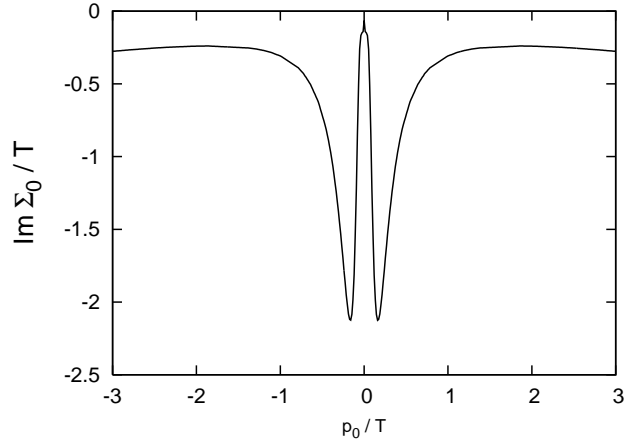}
      \caption{The imaginary part of the self-energy for $g=2$, $T/\Lambda=0.3$
      at rest ($p=0$).}
      \label{fig:ImSigma0}
\end{center}
\end{figure}
\begin{figure}[t]
\begin{center}
	\includegraphics[keepaspectratio,height=6cm]{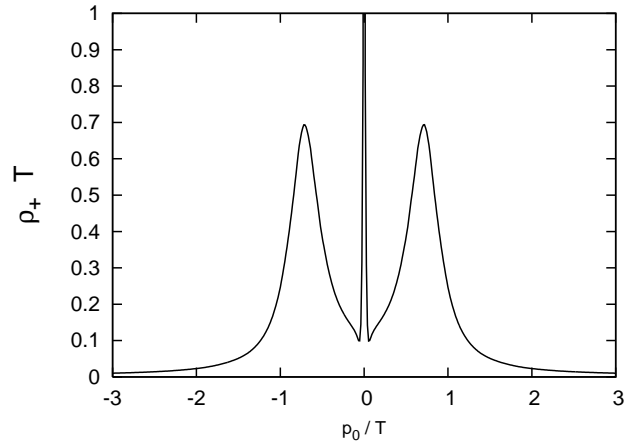}
      \caption{The spectral function for $g=2$, $T/\Lambda=0.3$ 
	at rest ($p=0$).}
      \label{fig:sp-HTL}
\end{center}
\end{figure}

Now, we present our numerical result for $g=2$. 
In Fig.~{\ref{fig:ImSigma0}} we plot the $p_0$ dependence of the imaginary
part of the self-energy, ${\rm Im} \Sigma_0(p_0,0)$.
We see two distinct peaks for $|p_0/M_{\rm HTL}|<1$, which come from the Landau
damping and the kinematics: 
The Landau damping implies a non-vanishing imaginary part that is even for 
$|p_0/M_{\rm HTL}|<1$, 
while the imaginary part vanishes at the origin, because of the suppression of
the decay rate for distribution functions. 
These two peaks in ${\rm Im} \Sigma_0(p_0,0)$
are similar to those in the system studied in Ref.~\citen{Kitazawa:2006zi}
with a massless fermion and a massive boson that result from a Yukawa coupling.
Because the current fermion mass is taken to be zero, 
the fermion spectrum is determined solely from the structure of the imaginary
part of the self-energy, as shown in Eq.~(\ref{eq:HTL2spc}). 
Therefore, the fermion spectrum should also be similar to that obtained in 
Refs.~\citen{Kitazawa:2006zi} and \citen{Mitsutani:2007gf}, 
and actually form a three-peak structure, as shown in Fig.~\ref{fig:sp-HTL}, 
which is qualitatively the same as that in
Refs.~\citen{Kitazawa:2006zi} and \citen{Mitsutani:2007gf}.

The peak around the origin in Fig.~\ref{fig:sp-HTL} is very sharp, because the
imaginary part of the self-energy vanishes at the origin, as mentioned above. 
Contrastingly, the peak at the origin obtained using the SDE is not so sharp, 
due to the non-perturbative iteration: 
If we use the spectral function (\ref{eq:HTL2spc}) for the internal fermion
line in Eq.~(\ref{eq:HTL-self}) instead of ${\mathcal S}_{\rm HTL}$ as an
effect of the iteration, the imaginary part of
the resultant self-energy will be non-zero even at the origin, 
and the peak around the origin should become broader.

\end{document}